\documentclass[]{pasj01}

\begin{document} 
\Received{2017/02/28}
\Accepted{}

\title{Discovery of H$_2$O Megamasers in Obscured Active Galactic Nuclei}

\author{Aya \textsc{Yamauchi},\altaffilmark{1,2} 
Yusuke \textsc{Miyamoto},\altaffilmark{3} 
Naomasa \textsc{Nakai},\altaffilmark{4,5} 
Yuichi \textsc{Terashima},\altaffilmark{6} 
Taishi \textsc{Okumura},\altaffilmark{4} Bin \textsc{Zhou},\altaffilmark{4} 
Kotomi \textsc{Taniguchi},\altaffilmark{3,7} 
Hiroyuki \textsc{Kaneko},\altaffilmark{3} 
Naoko \textsc{Matsumoto},\altaffilmark{8,9} 
Dragan \textsc{Salak},\altaffilmark{10} 
Atsushi \textsc{Nishimura},\altaffilmark{11} and Saeko \textsc{Ueno}\altaffilmark{12}}
\altaffiltext{1}{Research Fellow, Japan Society for the Promotion of Science}
\altaffiltext{2}{Mizusawa VLBI Observatory, National Astronomical Observatory of Japan, 2-12 Hoshigaoka, Mizusawa, Oshu, Iwate 023-0861, Japan}
\altaffiltext{3}{Nobeyama Radio Observatory, National Astronomical Observatory of Japan, 462-2
Nobeyama, Minamimaki, Minamisaku, Nagano 384-1305, Japan}
\altaffiltext{4}{Division of Physics, Faculty of Pure and Applied Sciences, University of Tsukuba, 1-1-1 Tennodai, Tsukuba, Ibaraki 305-8571, Japan}
\altaffiltext{5}{Center for Integrated Research in Fundamental Science and Engineering (CiRfSE), University of Tsukuba, Tennodai 1-1-1, Tsukuba, Ibaraki 305-8571, Japan}
\altaffiltext{6}{Department of Physics, Ehime University, Bunkyo-cho, Matsuyama, Ehime 790-8577, Japan}
\altaffiltext{7}{Department of Astronomical Science, School of Physical Science, SOKENDAI (The Graduate University for Advanced Studies), 2-21-1 Osawa, Mitaka, Tokyo 181-8588, Japan}
\altaffiltext{8}{The Research Institute for Time Studies, Yamaguchi University, 1677-1 Yoshida, Yamaguchi, Yamaguchi 753-8511, Japan}
\altaffiltext{9}{Mizusawa VLBI Observatory, National Astronomical Observatory of Japan, 2-21-1 Osawa, Mitaka, Tokyo 181-8588, Japan}
\altaffiltext{10}{Department of Physics, School of Science and Technology, Kwansei Gakuin University, Gakuen 2-1 Sanda, Hyogo 669-1337, Japan}
\altaffiltext{11}{Department of Physics, Nagoya University, Chikusa-ku, Nagoya 464-8602, Japan}
\altaffiltext{12}{Graduate School of Science and Engineering, Kagoshima University,
1-21-35 Korimoto, Kagoshima, Kagoshima 890-0065, Japan}
\email{a.yamauchi@nao.ac.jp}


\maketitle

\begin{abstract}
Recently a new method to discover obscured active galactic nuclei (AGNs) by utilizing X-ray and Infrared data has been developed.
We carried out a survey of H$_2$O maser emission toward ten obscured AGNs with the Nobeyama 45-m telescope.
We newly detected the maser emission with the signal-noise-ratio ({\it SNR}) of above 4 from two AGNs, NGC 1402 and NGC 7738.
We also found a tentative detection with its ${\it SNR} > 3$ in NGC 5037.
The detection rate of $20 \%$ is higher than those of previous surveys (usually several percents).
\end{abstract}

\section{Introduction}

\begin{table*}[htb!]
\caption{Galaxies searched for H$_2$O maser emission}
\begin{tabular}{lccccccc}
\hline
Name	& $\alpha_{2000.0}$ & $\delta_{2000.0}$ & $z$	& $N_{\rm H}$$^*$ & $L_{2-10}$$^*$ & Optical$^*$ & X-ray$^*$\\
	& (h m s) & ($\timeform{D}\ \timeform{'}\ \timeform{''}$) &	& ($10^{22}$ cm$^{-1}$) & ($10^{42}$ erg s$^{-1}$) & Class	& spectra \\
(1)	& (2)	& (3)	& (4)	& (5)	& (6)	& (7) 	& (8) \\
\hline
UGC 2730	& 03 25 52.340 & $+$40 44 55.60 & 0.0126	& $21^{+25}_{-14}$& 0.0481	& Unclass	& A \\
NGC 1402	& 03 39 30.570 & $-$18 31 37.00 & 0.0143	& $>100$	& 0.017	& H\emissiontype{II}/L	& F \\
NGC 2611	& 08 35 29.170 & $+$25 01 39.00 & 0.0175	& $140^{+150}_{-80}$	& 2.11 & L	& A \\
SDSS J085312$^{\dagger}$	& 08 53 12.354 & $+$16 26 19.51 & 0.0638	& $5.3^{+1.9}_{-1.5}$	& 3.50 & H\emissiontype{II}	& A \\
IC 614	& 10 26 51.850 & $-$03 27 53.10 & 0.0342	& $>100$	& 0.923 & Sy2	& F \\
CGCG 213-027	& 11 05 21.897 & $+$38 14 01.83 & 0.0285	& $34^{+30}_{-17}$	& 0.764 & L	& A \\
NGC 5037	& 13 14 59.370 & $-$16 35 25.10 & 0.0099	& $48^{+25}_{-19}$	& 0.075 & Unclass 	& A \\
UGC 8621	& 13 37 39.870 & $+$39 09 16.99 & 0.0201	& $54^{+59}_{-28}$	& 0.376 & Sy2	& A \\
NGC 5689	& 14 35 29.694 & $+$48 44 29.95 & 0.0072	& $27^{+19}_{-12}$	& 0.079	& normal	& A \\
NGC 7738 	& 23 44 02.059 & $+$00 30 59.86 & 0.0226	& $>100$	& 0.042& H\emissiontype{II}/L	& F \\
\hline
\end{tabular}
\label{target}
\begin{tabnote}
(1) Galaxy name ($\dagger$SDSS J085312.35+162619.4), (2) right ascension and (3) declination of the center position from NED, (4) redshift from NED, (5) absorption column density, (6) absorption-corrected luminosity in 2--10 keV in units of $10^{42}$ erg s$^{-1}$, (7) optical classification; Sy2-Seyfert 2, L-LINER, H\emissiontype{II}-H\emissiontype{II} nucleus, normal-normal galaxy, Unclass-unclassified, and (8) X-ray spectra; A-absorbed, F-flat continuum and strong Fe-K line. \\
$*$ \citet{tera15}, except for UGC 2730 (private communication).
\end{tabnote}
\end{table*}

More than 160 extragalactic H$_2$O masers have been discovered, mainly from active galactic nuclei (AGN)\footnotemark.
In the active galaxy NGC 4258, H$_2$O maser features around its systemic velocity and extremely high-velocity features symmetrically offset from the systemic velocity by $\pm 1000$ km s$^{-1}$ were detected (\cite{nakai93}).
Very long baseline interferometry observations of the maser emission have revealed the existence of a compact disk in a Keplerian rotation and of a massive black hole at its nucleus (\cite{miyo95}).
In this way, H$_2$O maser emission is a unique probe to directly investigate the structure and dynamics of AGNs on the (sub-)parsec scale.
\footnotetext{The data available at The Megamaser Cosmology Project website: https:{\slash\slash}safe.nrao.edu/wiki/bin/view/Main/MegamaserCosmologyProject.}

Hard X-ray above 2 keV investigation provides a helpful tool to identify nuclear activity, because hard X-ray can penetrate a large amount of absorbing material (e.g., \cite{corral14}). In case that the hydrogen column density $N_{\rm H}$ is larger than $10^{24}$  cm$^{-2}$ (``Compton-thick"), however, even hard X-ray is heavily absorbed and the strong iron (Fe) K line is seen (\cite{coma05}).

Infrared (IR) re-emission from dust illuminated by AGNs is also useful to find obscured AGNs  (\cite{ichi12}). A weakness of utilizing infrared emission, however, is that the emission from dust heated by Seyfert class AGNs cannot be distinguished from emission from dust heated by stars, if spatial resolution is not sufficiently high using ground-based telescopes with narrow-slit spectroscopy.

Combination of X-ray and IR is the most efficient tool to find heavily obscured AGNs, because it can overcome the selection biases utilizing only one of X-ray spectra and IR emission.
\citet{tera15} newly found 48 obscured AGNs, cross-correlating 18 and 90 $\mu$m sources in the AKARI point-source catalogs (\cite{IRC}, \cite{FIS}) and X-ray sources in the XMM-Newton serendipitous source catalog (\cite{XMM}).
Many of previous H$_2$O maser surveys have targeted AGNs classified by optical observations. In this paper, we adopted the obscured AGNs as our sample to find new H$_2$O masers in AGNs.

\section{Observations}

We observed ten obscured AGNs found by cross-correlating XMM-Newton and AKARI catalogs in H$_2$O maser emission ($J_\mathrm{K_{a}K_{c}} = 6_{16}$--$5_{23}$ rotational transition at 22.23508 GHz).
Nine objects, for which clear evidence for the presence of obscured AGNs is seen in their X-ray spectra, were taken from \citet{tera15}. 
One additional object, UGC 2730, was selected in the same manner, but its X-ray counts measured with EPIC-PN in the 0.2--12 keV band is slightly lower than the threshold applied in \citet{tera15}. 
Table \ref{target} shows the absorption column density ($N_{\rm H}$), X-ray luminosities, and classifications of X-ray spectra (``Absorbed" or ``Flat") for the obscured AGN component of the objects in our sample. 
Optical classifications are also summarized in table \ref{target}; our sample contains galaxies that are not previously classified as AGNs from optical observations. 
The velocity in this paper is defined with respect to the local standard of rest (LSR) in radio definition.

We made observations of H$_2$O maser emission from February to June in 2016 using the 45-m telescope of the Nobeyama Radio Observatory\footnotemark (NRO). 
We additionally observed IC 614 only in December 2016.
The half-power beam width and the aperture efficiency of the telescope were $\mathrm{HPBW} = \timeform{74''} \pm \timeform{1''}$ and $\eta_\mathrm{a} = 0.61 \pm 0.03$ at 22 GHz, respectively.
\footnotetext{Nobeyama Radio Observatory is a branch of the National Astronomical Observatory of Japan, National Institutes of Natural Sciences.}

The front-end receivers utilized HEMT amplifiers cooled to 20 K, equipped with two polarized feeds that received right and left-circular polarization simultaneously. 
The system noise temperature, $T_\mathrm{sys}$, including the atmospheric effect and the antenna ohmic loss, was 98--529 K, depending on the weather conditions and the observing elevations of $\timeform{20D}$--$\timeform{80D}$. 

The receiver back-ends were SAM45 (Spectral Analysis Machine for the 45-m telescope, \cite{kama12}). 
We used eight sub-bands of the 125-MHz bandwidth each with 4096 spectral points providing the 30.52-kHz frequency resolution.
The corresponding velocity coverage and velocity resolution were 1685 km s$^{-1}$ and 0.41 km s$^{-1}$ at 22 GHz, respectively. 
A center frequency of each sub-band was 22.1725, 22.2350, or 22.2975 GHz, resulting in a total frequency coverage of 22.1100--22.3600 GHz (a total velocity coverage of 3370 km s$^{-1}$).

The observations were made in the position-switching mode with an off-source position of $\timeform{7'}$ in the azimuth direction. 
The pointing accuracy was about $\timeform{10''}$. 
The calibration of the line intensity was performed by chopping the sky and a reference load at room temperature, yielding an antenna temperature, $T^{*}_\mathrm{A}$, corrected for the atmospheric attenuation. 
The measured $T^{*}_\mathrm{A}$ was converted into the flux density, $S$, using the sensitivity, $S/T^{*}_\mathrm{A} = 2.85 \pm 0.14$ Jy K$^{-1}$, calculated from the aperture efficiency.

The data reductions were made by using the NEWSTAR software package developed by NRO.
After flagging bad scan data in each observing epoch, we integrated the data with a two-channel binning mode. 
Next, the best-fit linear gradient was subtracted to determine a baseline of each spectrum. 
Then, we averaged the spectra over all observing epochs. 
Finally, all the spectra were binned to velocity resolutions of 1 and 5 km s$^{-1}$.
Table \ref{status} lists observing status.

The integration time of each observation epoch was too short to detect the signal.
Therefore, it is difficult to know whether or not the flux density and velocity of the maser features vary with time using our data.

\begin{table}[htb!]
\caption{Observation status.}
\begin{tabular}{lccc}
\hline
Name	& $V_{\rm sys}$$^*$ & $V_{\rm obs}$$^{\dagger}$ & rms$^{\ddagger}$ \\
	& (km s$^{-1}$) & (km s$^{-1}$) & (mJy) \\
\hline
UGC 2730& 3721 & 2086--5498 & 13 \\
NGC 1402& 4220 & 2403--5819 & 9 \\
NGC 2611& 5158 & 3432--6860 & 10 \\
SDSS J085312	& 17968&16159--19743& 11 \\
IC 614	& 9911 & 8223--11708& 8  \\
CGCG 213-027	& 8303 & 6548--10014& 11 \\
NGC 5037& 1848 &  184--3574 & 11 \\
UGC 8621& 5916 & 4194--7632 & 9 \\
NGC 5689& 2159 &  549--3943 & 11 \\
NGC 7738& 6614 & 5021--8467 & 11 \\
\hline
\end{tabular}
\label{status}
\begin{tabnote}
$*$ Velocity converted from $V_{\rm opt, helio}$ in NED except \citet{fou92} for NGC 5037.\\
$\dagger$ Velocity range searched for H$_2$O maser emission.\\
$\ddagger$ Rms of a spectrum with its velocity resolution of 1 km s$^{-1}$.
\end{tabnote}
\end{table}

\section{Results and Discussion}

\begin{figure}
 \begin{center}
  \includegraphics[width=8cm]{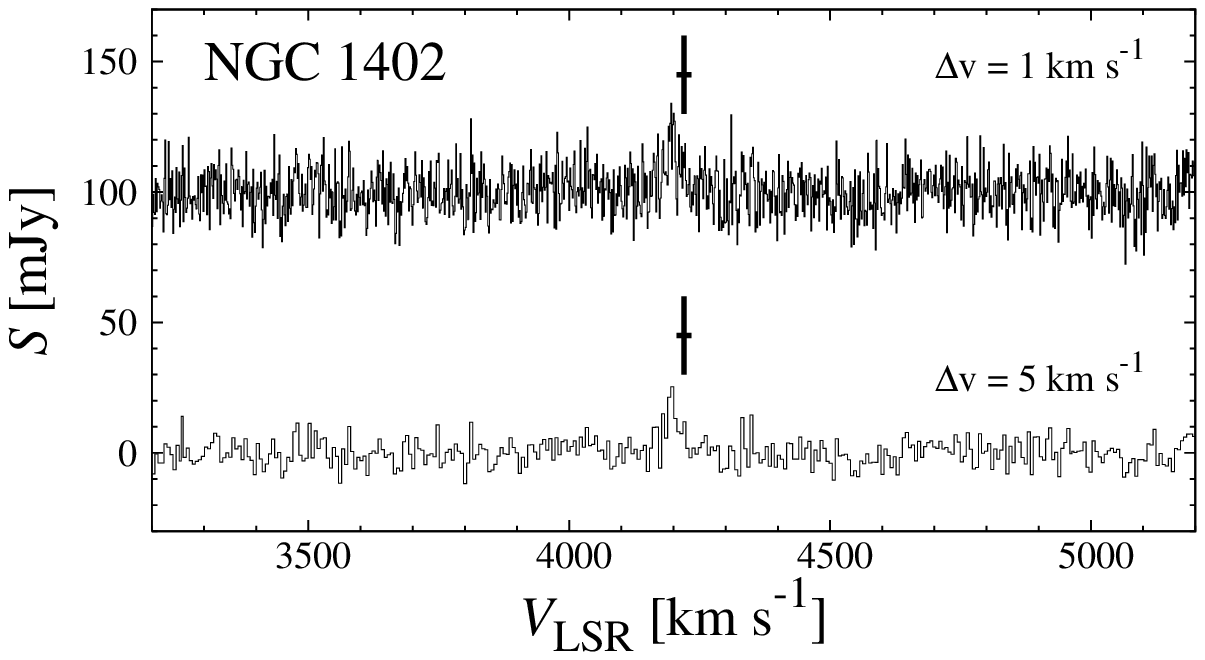}
  \includegraphics[width=8cm]{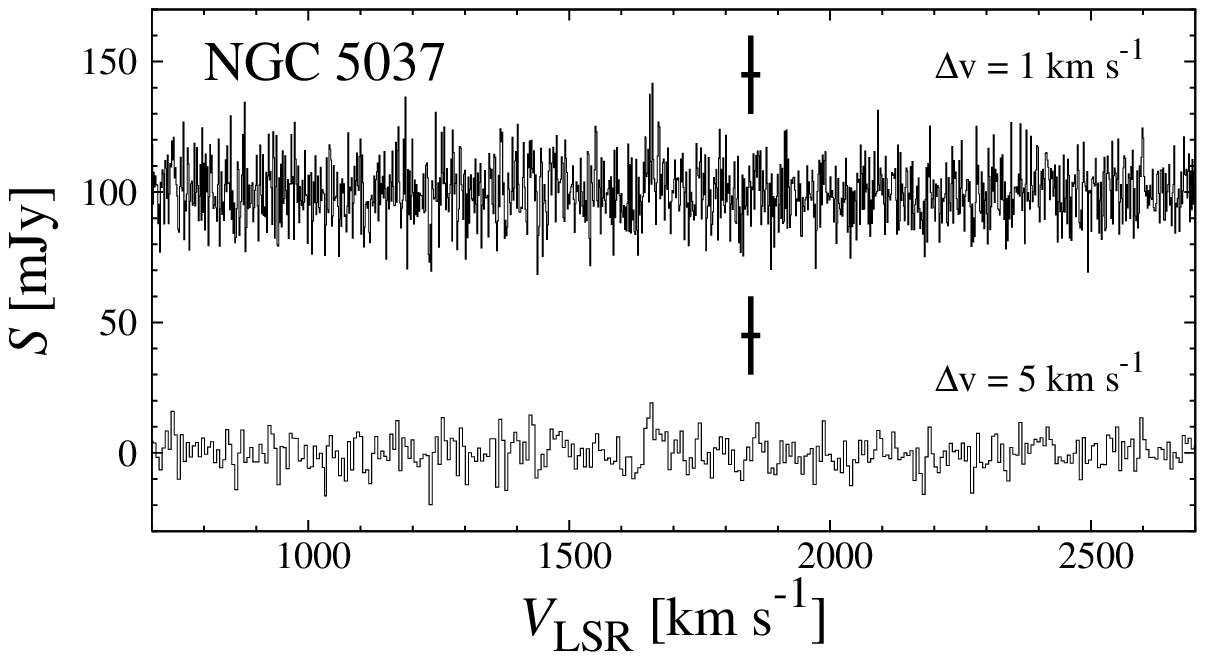}
  \includegraphics[width=8cm]{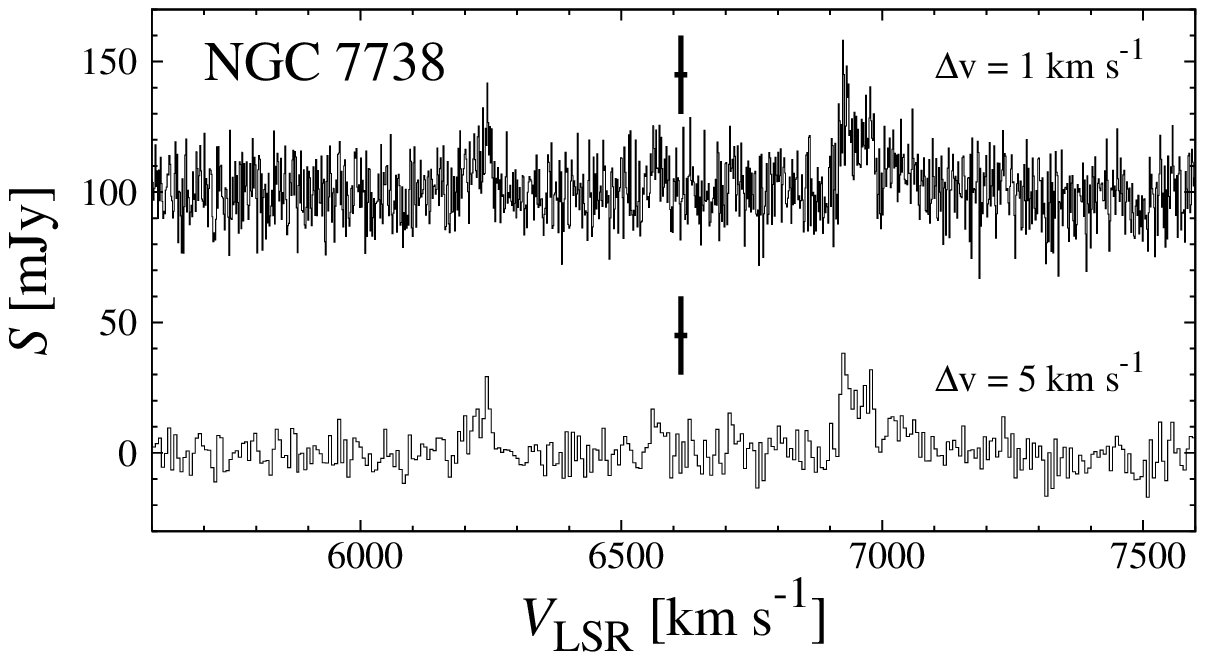}
 \end{center}
\caption{The H$_2$O maser spectra of AGNs (marginally) detected with the 45-m telescope. 
To make the graph easier to view, the spectra with the velocity resolution of 1 km s$^{-1}$ are plotted 100 mJy above.
Vertical and horizontal lines in the spectra indicate the systemic velocities of the galaxies and those errors, respectively.}
\label{sp1}
\end{figure}

We newly detected maser features from two AGNs, NGC 1402 and NGC 7738, with the signal-to-noise-ratio ({\it SNR}) of above 4, and tentatively detected in NGC 5037 (${\it SNR} > 3$).
Figure \ref{sp1} shows the maser spectra of the AGNs, and table \ref{result} lists parameters of the masers.

NGC 1402 shows maser features with its peak velocity, 4195 km s$^{-1}$, at the systemic velocity [4220$\pm$14 km s$^{-1}$, NED reffers to \citet{firth14} using the optical redshift] of the galaxy (figure \ref{sp1}), indicating the systemic features. The isotropic maser luminosity of 47 $\LO$ is that of a megamaser (i.e., $\geq 10\ \LO$).

NGC 7738 shows maser features redshifted and blueshifted from the systemic velocity [6614$\pm$12 km s$^{-1}$, NED reffers to \citet{theu98} using the H\emissiontype{I} 21-cm line] of the galaxy (figure \ref{sp1}). Weak features at $V_{\rm LSR} \approx 6573$ km s$^{-1}$ may be systemic velocity features, although {\it SNR} is not enough.
Redshifted and blueshifted features are symmetrical with respect to the possible systemic features in velocity. Such symmetrical three components are typically seen as water-vapor masers in AGNs (e.g., \cite{nakai93}, 1995, \cite{green09}), indicating a rotating edge-on disk (its inclination angle is $\approx \timeform{90D}$; e.g., \cite{miyo95}, The Megamaser Cosmology Project).
Thus the triple peak-like spectrum in NGC 7738 strongly suggest an edge-on maser disk rotating with velocity of $\sim 350$ km s$^{-1}$ (separation between the systemic and high velocity features), with a quite different angle from the galactic disk whose inclination angle is $\sim \timeform{60D}$ (NED), as seen in many megamasers in AGNs (e.g., \cite{miyo95}, \cite{yama12}, \cite{green09}).
The isotropic maser luminosity of 468 $\LO$ is that of a megamaser.

NGC 5037 shows possible maser features at $V_{\rm LSR} = 1660$ km s$^{-1}$ maybe offset from the systemic velocity [1848$\pm$18 km s$^{-1}$, \citet{fou92} using the optical redshift] of the galaxy by $\sim 190$ km s$^{-1}$. The maser luminosity of $L_{\rm iso} = 5\ \LO$ is one order of magnitude smaller than typical luminosities of megamasers. More sensitive observations are required to confirm the maser features.

NGC 1402 and NGC 7738 have the X-ray spectra with a flat continuum and a strong Fe-K emission line, indicating the presence of Compton-thick AGNs in these objects.
From NGC 6926 classified as an AGN of the same type by \citet{tera15}, H$_2$O maser emission has already been detected (\cite{green03}, \cite{sato05}).
This type of AGNs may tend to have H$_2$O maser emission.

\begin{table}[htb!]
\caption{Detected maser features with 1 km s$^{-1}$ resolution.}
\begin{tabular}{lcccc}
\hline
Name	& $V_{\rm peak}$& $S_{\rm peak}$& {\it SNR} 	& $L_{\rm iso}$$^*$ \\
	& (km s$^{-1}$)	& (mJy)		& ($\sigma$)	& ($\LO$) \\
\hline
NGC 1402& 4195 & 34 & 4.0 	& 47 \\
NGC 5037& 1660 & 42 & 3.9 	& 5 \\
NGC 7738& 6243 & 42 & 4.0 	& 131 \\
	&(6573 & 26 & 2.4	& 75)$^{\dagger}$ \\
	& 6925 & 58 & 5.5 	& 337 \\
	& total&    &		& 468 \\
	&      &    &		& (543)$^{\dagger}$ \\
\hline
\end{tabular}
\label{result}
\begin{tabnote}
$*$ Isotropic luminosity $L_{\rm iso}\ [\LO] = 0.023 \times \int S dV\ [{\rm Jy\ km\ s}^{-1}] \times (D\ [\rm Mpc])^2 $. The distances, $D$, to NGC 1402, NGC 5037, and NGC 7738 are 58 Mpc, 29 Mpc, and 94 Mpc, respectively, adopted from NED (NGC 1402, NGC 7738) and \citet{tully88} (NGC 5037). \\
$\dagger$ Possible systemic velocity features and the total luminosity including the systemic features.
\end{tabnote}
\end{table}

\citet{sato05} made a survey for AGNs with the velocity of $cz < 10000$ km s$^{-1}$, using the NRO 45-m telescope. Their targets were Seyfert 1.8 to 2 or LINER.
The detection rate was $\sim 1 \%$ ($1/93$).
\citet{green03} observed AGNs with $cz < 8100$ km s$^{-1}$ with the 70-m antenna of the NASA Deep Space Network (their sample selection source was not described). Their detection rate was $\sim 4\%$ ($7/160$). 
The Megamaser Cosmology Project reported a large survey of over 3000 AGNs using the Robert C. Byrd Green Bank Telescope and the detection rate of $\sim 3\%$. Survey candidates included not only Seyfert 2 and LINER but also AGNs identified by Swift/BAT (\cite{braatz15}; We expect the details of sample selection in their forthcoming paper.). 
Hagiwara \etal\ (2002, 2003) selected their targets based on the ratio of radio continuum flux density to IR (60 $\mu$m and 100 $\mu$m) flux density from IRAS galaxies and observed 24 AGNs with the Max-Planck-Institut f\"{u}r Radioastronomie 100-m telescope, resulting in the detection rate of $\sim 8\%$ ($2/24$).
Compared to these results, our detection rate of $20\%$ ($2/10$) is very high, although the number of the detected samples is statistically not enough.
This result may indicate the availability of the selection method combined X-ray with IR.

Through X-ray data, \citet{green08} and \citet{cast13} analyzed the column density of active galaxies whose H$_2$O maser emission has already been detected.
\citet{cast13} reported that 27 out of 47 samples were Compton-thick AGNs with $N_{\rm H} > 10^{24}$  cm$^{-2}$.
Limiting the sample to disk-like maser only, the percentage of Compton-thick AGNs increases from 57\% to 78\% with 18/23. 
\citet{masini16} studied the connection between the torus (seen as the X-ray obscurer) and the maser disk.
They proposed a toy model that the maser disk is the inner part of the torus, ending in an inflated, geometrically thicker structure.
This model was able to recover the column density distribution for a sample of AGNs and explains the connection between H$_2$O megamaser emission and nuclear obscuration as a geometric effect. They said that a more physical picture explicitly addressing the known disk/torus clumpiness and warping must rely on numerical calculations.
Further studies would be required to make clear the relationship between H$_2$O maser detection and Compton-thick AGNs.

\section{Summary}

We surveyed H$_2$O maser emission for ten AGNs and newly detected from two AGNs.
The targets galaxies were selected from obscured AGNs found by \citet{tera15} using X-ray and IR.
Our high detection rate indicates that the selection method is effective.
Both of NGC 1402 and NGC 7738 with maser detection are classified as Compton-thick AGNs from their X-ray spectra, suggesting a high detection rate of maser emission from this class of AGNs.

\begin{ack}
We thank members of NRO for their continuous support.
This research has made use of the NASA/IPAC Extragalactic Database (NED) which is operated by the Jet Propulsion Laboratory, California Institute of Technology, under contract with the National Aeronautics and Space Administration.
\end{ack}


\end{document}